\newcommand*{\wn}{cm$^{-1}$}
\newcommand{\etal}{\emph{et al.}}
\renewcommand{\thefootnote}{\fnsymbol{footnote}}

\def\pr{Phys.\  Rev.\ }

\def\cpl{Chem.\ Phys.\ Lett.\ }

\def\jms{J. Mol.\ Spectrosc.\ }
\def\jpb{J. Phys.\ B }
\def\rsi{Rev. Scient.\ Instr. }

\def\cjp{Can.\ J. Phys.\ }

\documentclass[aps,prb,twocolumn,superscriptaddress,showpacs,floatfix]{revtex4}
\usepackage{graphics}
\usepackage{longtable}
\usepackage{rotating}

\begin{document}

\title{Synchrotron VUV radiation studies of the D$ ^{1}\Pi_{u}$ State of H$_{2}$}

\author{G.D. Dickenson}
\affiliation{Institute for Lasers, Life and Biophotonics Amsterdam,
VU University, De Boelelaan 1081, 1081 HV  Amsterdam, The Netherlands}
\author{T.I. Ivanov}
\affiliation{Institute for Lasers, Life and Biophotonics Amsterdam,
VU University, De Boelelaan 1081, 1081 HV  Amsterdam, The Netherlands}
\author{M. Roudjane \footnote[8]{Research Associate at Laboratoire d'\'Etude du Rayonnement et de la Mati\`ere
en Astrophysique, UMR 8112 du CNRS, Observatoire de Paris-Meudon, 5
place Jules Janssen, 92195 Meudon cedex, France} \footnote[2]{Current Address: Department of Chemistry, 009 Chemistry-Physics Building, University of Kentucky, 505 Rose Street, Lexington, KY 40506-0055,USA}} 
\affiliation{Synchrotron Soleil, Orme des Merisiers, St Aubin BP 48, 91192 GIF sur Yvette cedex, France}
\author{N. de Oliveira}
\affiliation{Synchrotron Soleil, Orme des Merisiers, St Aubin BP 48, 91192 GIF sur Yvette cedex, France}
\author{D. Joyeux}
\affiliation{Synchrotron Soleil, Orme des Merisiers, St Aubin BP 48, 91192 GIF sur Yvette cedex, France}
\affiliation{Laboratoire Charles Fabry de l'Institut d'Optique, CNRS, Univ Paris-Sud Campus Polytechnique, RD128, 91127, Palaiseau cedex, France}
\author{L. Nahon}
\affiliation{Synchrotron Soleil, Orme des Merisiers, St Aubin BP 48, 91192 GIF sur Yvette cedex, France}
\author{W.-\"{U} L. Tchang-Brillet}
\affiliation{Laboratoire d'\'Etude du Rayonnement et de la Mati\`ere
en Astrophysique, UMR 8112 du CNRS, Observatoire de Paris-Meudon Universit\'e UPMC, 5
place Jules Janssen, 92195 Meudon cedex, France}
\author{M. Glass-Maujean}
\affiliation{Laboratoire de Physique Mol\'eculaire pour l'Atmosph\`ere et l'Astrophysique UMR7092, Universit\'e P. et M. Curie, 4 place Jussieu 75252 Paris cedex 05 France}
\author{I. Haar}
\affiliation{Institute of Physics and Center for Interdisciplinary Nanostructure Science and Technology, Heinrich Plett Str 40, 34132 Kassel, Germany}
\author{A. Ehresmann}
\affiliation{Institute of Physics and Center for Interdisciplinary Nanostructure Science and Technology, Heinrich Plett Str 40, 34132 Kassel, Germany}
\author{W. Ubachs\footnote[3]{Corresponding author. Email: wimu@few.vu.nl}}
\affiliation{Institute for Lasers, Life and Biophotonics Amsterdam,
VU University, De Boelelaan 1081, 1081 HV  Amsterdam, The Netherlands}

\date{\today}

\begin{abstract}
The $3p\pi D^{1}\Pi_{u}$ state of the H$_{2}$ molecule was reinvestigated with different techniques at two synchrotron installations. The Fourier-Transform spectrometer in the vacuum ultraviolet wavelength range of the DESIRS beamline at the SOLEIL synchrotron was used for recording absorption spectra of the $D^{1}\Pi_{u}$ state at high resolution and high absolute accuracy, limited only by the Doppler contribution at 100 K. From these measurements line positions were extracted, in particular for the narrow resonances involving $^{1}\Pi_{u}^-$ states, with an accuracy estimated at 0.06 \wn . The new data also closely match MQDT-calculations performed for the $\Pi^{-}$ components observed via the narrow Q-lines. The $\Lambda$-doubling in the $D^{1}\Pi_{u}$ state was determined up to $v=17$. The 10 m normal incidence scanning monochromator at the beamline U125/2 of the BESSY II synchrotron, combined with a home built target chamber and equipped with a variety of detectors was used to unravel information on ionization, dissociation and intramolecular fluorescence decay for the $D^{1}\Pi_{u}$ vibrational series. The combined results yield accurate information of the characteristic Beutler-Fano profiles associated with the strongly predissociated $\Pi_{u}^+$ parity components of the $D^{1}\Pi_{u}$-levels. Values for the parameters describing the predissociation width as well as the Fano-$q$ line shape parameters for the $J=1$ and $J=2$ rotational states were determined for the sequence of vibrational quantum numbers up to $v=17$.

\end{abstract}

\maketitle

\section{Introduction}

\begin{figure}[pt]
\begin{center}
\includegraphics[width=1\linewidth]{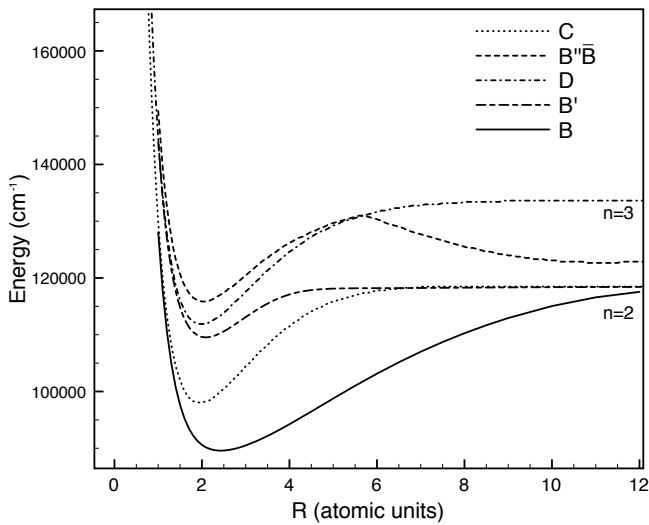}
\caption[]{The potential energy curves of the relevant excited electronic states~\cite{Dressler1986}.}
\label{fig:PotentialEnergyCurves}
\end{center}
\end{figure}

The energy region from 110 000 to 134 000 \wn\ in the absorption spectrum of molecular hydrogen is dense and complex featuring a multi-line spectrum. In particular in this energy range the electronic and vibrational energy separations become of similar magnitude, while rotational splittings
remain large as well.
The $D^{1}\Pi_{u} - X^{1}\Sigma_{g}^{+}$ system gives rise to a well-marked sequence of intense resonances in this region. Hopfield observed the $D^{1}\Pi_{u}$ state for the first time\cite{Hopfield1930}, but failed to find the correct sequence of vibrational levels. This was established in the early work by Beutler and coworkers\cite{Beutler1935a} and by Richardson\cite{Richardson1935}. From an extrapolation of the vibrational sequence they deduced that below the lowest predissociated level there exist three lower vibrational levels $v=0-2$, with weaker apparent intensity. These assignments of the vibrational sequence in the $D^{1}\Pi_{u}$ state was later confirmed in subsequent higher resolution studies by Namioka\cite{Namioka1964}, Monfils\cite{Monfils1965}, Takezawa\cite{Takezawa1970} and Herzberg and Jungen\cite{Herzberg1972}. A potential energy curve of the $D^{1}\Pi_{u}$ state, calculated by Dressler and Wolniewicz~\cite{Dressler1986} along with the other relevant potential curves of singlet-$u$ symmetry for interpreting the results of the present study is displayed in Fig.~\ref{fig:PotentialEnergyCurves}. 

The unpredissociated states, giving rise to sharp lines in the spectrum, have  been studied to a high degree of accuracy. Analysis of classical spectrograph emission spectra by Abgrall \etal\cite{Abgrall1994} yielded accurate level energies up to high rotational quantum numbers for both parity components of $D^{1}\Pi_{u}\,(v=0-2)$. Since the $\Pi^-$ parity components of the higher lying vibrational levels $v>2$ of the $D^{1}\Pi_{u}$ state undergo only very weak predissociation~\cite{Guyon1979,Glass-Maujean1985}, the Q lines probing $D^{1}\Pi^{-}_{u}$ levels are observed as sharp. That made it possible for Abgrall~\etal~\cite{Abgrall1994} to also determine accurate line positions from emission studies for lines probing up to $v=14$.
Extreme ultraviolet laser-based absorption was used by Reinhold \etal\cite{Reinhold1996} to record high-resolution spectra of the $D-X(1,0)$ band at even better accuracy. The most accurate level energies of the unpredissociated levels were obtained in the work combining Doppler-free laser excitation and visible and near-infrared Fourier-Transform emission spectroscopy\cite{Hannemann2006,Bailly2010} yielding accuracies as good as $0.005$ \wn\ for $ D $ $v=0-2$  levels.

Beutler \cite{Beutler1935a} reported on the observation of asymmetrically broadened line shapes in the R-branches of the $D-X$ system for excited vibrational levels $v > 2$. These shapes were interpreted as Beutler-Fano profiles, as resulting from an interference between a bound and a continuum state in an excitation spectrum, as first observed in the autoionization of noble gas atoms by Beutler\cite{Beutler1935b} and explained by Fano\cite{Fano1935,Fano1961}. The first  theoretical calculations of the widths were performed ~\cite{Julienne1971,Fiquet-Fayard1971,Fiquet-Fayard1972} in 1971, simultaneously with an experimental study by Comes and Schumpe \cite{Comes1971}. It was settled that the predissociation of the $ \Pi^{+} $ parity component of the $D^1\Pi_u$ state must be attributed mainly to an interaction with the $B'^1\Sigma^+_u$ continuum. A more recent calculation~\cite{Beswick1988} shows that the $ 2p\sigma B^{1}\Sigma^{+}_{u} $ and $ 2p\pi C\Pi^{+}_{u} $ continua have only an effect on the ratio between H(2s) and H(2p) dissociation fragments. Mental and Gentieu\cite{Mentall1970} and later Guyon~\etal\cite{Guyon1979} performed measurements detecting Ly$-\alpha$ radiation emitted by the dissociation product. A pioneering study was conducted by Jungen and Atabek~\cite{Jungen1977} in which a full MQDT treatement of the $B,B',C,B''$ and $D$ systems was conducted. Level energies up to $v=13$ were calculated for the $\Pi^{-}$ components. The $\Lambda$-doubling was determined for the unpredissociated  bands (0,0)-(2,0) for levels $J$=1,3 and 5 as well as for the first predissociated band (3,0) for $J$=1.

The issue of the asymmetry of the line shapes and the Fano-$q$ parameters was studied in detail by Glass-Maujean~\etal~\cite{Glass-Maujean1979} and by Dehmer and Chupka\cite{Dehmer1980}. 

In addition to absorption with classical light sources also lasers were used to
investigate the predissociation in the $D^{1}\Pi_{u}$ state. The studies by Rothschild \etal~\cite{Rothschild1981} used a laser with a superior instrument width of $0.005$ \wn\ to investigate H$_2$ at room-temperature Doppler broadening exceeding $1$ \wn\ to yield accurate widths and asymmetry paramaters for a number of lines, mainly of hydrogen isotopomers. Croman and McCormack~\cite{Croman2008} performed two-step laser-excitation investigating the $v=12$ and $13$ states of $D^1\Pi_u$.

The present spectroscopic study reinvestigating the $D^{1}\Pi_{u}-X^1\Sigma_g^+$ system in H$_2$ is based on two different experimental approaches, both connected to a major synchrotron facility. The novel and unique Fourier-Transform spectrometer at the SOLEIL synchrotron facility (France) on the DESIRS VUV beamline used to record Fourier Transform (FT) absorption spectra in the relevant range $74-90$ nm from a static gas sample of H$_2$ cooled to approximately 100 K, resulting in high resolution spectra, limited by the Doppler contribution of 0.6 \wn, while the FT instrument resolution is 0.35 \wn. Apart from a determination of accurate transition frequencies for the narrow unpredissociated resonances the asymmetrically broadened resonances have been studied at high resolution to obtain information on the Fano $q$ parameters as well as predissociation widths $\Gamma$ for individual lines. The 10 m normal - incidence monochromator at beamline U125/2 of the BESSY II synchrotron radiation facility in Berlin combined with a home-built target chamber for measuring photoabsorption, photoionization and photodissociation by means of fragment fluorescence \cite{Glass-Maujean2007,Glass-Maujean2010} was used to compare the different decay channels upon excitation of the same $D^{1}\Pi_{u}-X^1\Sigma_g^+$ system. The latter spectra recorded at slightly lower resolution aided in assigning and disentangling the observed features.

\section{Experimental setup}

\begin{figure}[pt]
\begin{center}
\includegraphics[width=1\linewidth]{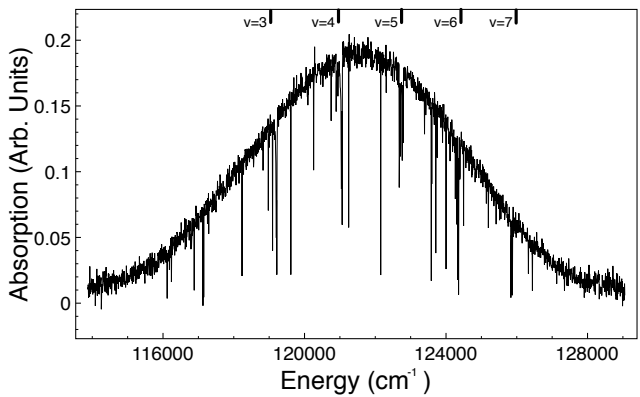}
\caption[]{Spectral recording of an absorption spectrum of H$_2$ for a single
setting of the undulator optimum. Some of the $D-X\,(v,0)$
resonances are indicated.}
\label{fig:Bell-curve}
\end{center}
\end{figure}

The major part of the experimental data in the present study was obtained with the vacuum ultraviolet Fourier-Transform spectrometer setup, connected to the DESIRS beamline at the SOLEIL synchrotron. Its principle of operation and unique capabilities for the UV and VUV range are described elsewhere \cite{deOliveira2009,deOliveira2010}. In short, a scanning wave-front division interferometer has been specifically developed in order to extend the FT spectroscopy technique toward the VUV range \cite{deOliveira2009}. The undulator based DESIRS beamline provides the 7 $\%$ bandwidth continuum background which is analyzed by the FT-spectrometer after it has passed an absorption cell. A typical recording of such a full spectrum recorded after passing the absorption cell is displayed in Fig.~\ref{fig:Bell-curve}.  Recently the same setup was used to record high resolution spectra of narrow transitions in the Lyman and Werner bands of the HD molecule in the wavelength range $87-112$ nm\cite{Ivanov2010}. The undulator profile covering roughly 5000 \wn\ is well approximated by a Gaussian function. The spectral range from 115 000 - 135 000 \wn, covered in the present study, was divided into four overlapping measurement windows.

The H$_2$ spectra are recorded under quasi-static conditions. Hydrogen gas passes through a needle valve regulator into a T-shaped, windowless free flow cell of 100 mm length and 12 mm inside diameter. This results in an inhomogeneous density distribution along the path of the synchrotron radiation, for which an integrated column density can only be estimated. It is set by regulating the pressure just outside the target cell for appropriate absorption conditions, \emph{i.e} unity optical depth for the main features under investigation.
Spectra were recorded at three different settings of the pressure at the inlet needle valve.

Once the radiation has passed through the target cell it enters the spectrometer and an interferogram is generated. For the present study recordings are taken at 512 ks, that is the number of sampling points taken along the path of the moving arm in the interferometer traveling over $\sim 9$ mm. A typical measurement window in the present study took roughly two hours for accumulating signal over 100 interferograms. The setting of 512 ks sampling recording corresponds to a spectral resolution of 0.35 \wn\ or a resolving power of 350 000.

The target cell was enveloped by a second cell in which liquid nitrogen is allowed to flow, cooling the cell down to roughly 100 K. This was done for two reasons. At the lower temperatures the complexity of the spectrum and the overlap of resonances is reduced; only the lowest rotational states up to $J=2$ were found to be populated. At this temperature the Doppler width of the H$_2$ lines reduces to $0.6$ \wn . With the chosen setting of the resolving power of the FT-spectrometer this yields the narrowest width for the unpredissociated lines at $\sim 0.7$ \wn.

An advantage of an FT-spectrometer is that the wavelength is intrinsically calibrated, when the travel distance in the moving arm is known; this is done by referencing the travel against the fringes of a frequency stabilised HeNe laser\cite{deOliveira2009}. Additional calibration of the frequency scale is accomplished
by calibration against the Ar line\cite{Sommavilla2002} $(3p)^{5}(^{2}\mathrm{P}_{3/2})9d[3/2] \leftarrow (3p)^{6}\, ^{1}\mathrm{S}_{0}$ known to an accuracy of 0.03 \wn.

The spectra recorded at the BESSY II Synchrotron in Berlin were measured with a 10 m normal incidence monochromator equipped with a 1200 lines/mm grating with a spectral resolution of 2 \wn\ or 0.0012 nm\cite{Glass-Maujean2007a}. The absorption cell is a 39 mm long differentially pumped cell containing 27 $\mu$bars of H$ _{2}$. A photodiode at the back of the cell allows for detecting direct absorption at room temperature. An electrode in the target cell attracts photoions and the Ly$-\alpha$ fluorescence is measured with a microchannel plate detector. Molecular fluorescence is recorded via a detector sensitive to visible radiation. All the signals mentioned above were recorded as a function of incident photon energy, so that reliable comparisons can be made and relate the cross sections of the various decay channels. This experimental setup allows absolute intensity measurements and quantitative dynamical studies.

\section{Predissociation in the $D^1\Pi_u$ state}

It was already understood at an early stage that predissociation of the $3p\pi D^{1}\Pi_{u}$ state
proceeds via coupling to the continuum of the $3p\sigma B'^{1}\Sigma^{+}_{u}$ state \cite{Julienne1971,Fiquet-Fayard1971}. There is no coupling between states of $u$ and $g$ inversion symmetry thus there are three candidates with the same $u$ symmetry for the $D$ state predissociation mechanism: the $2p\pi\,C^{1}\Pi_u$, $2p\sigma\,B^{1}\Sigma_u^{+}$ and $3p\sigma B'^{1}\Sigma_u^{+}$ states. The Coriolis coupling to the $B'$ state is the chief cause of the predissociation. Coupling to the $C$ state is two orders of magnitude weaker and coupling to the $B$ state is even smaller by at least one order of magnitude \cite{Gao1993}.  The $B'$ and $D$ states approach the same electronic He$(3p)$ configuration in the united atom limit so the Coriolis coupling may be very effective. The same interaction between $ D $ and $ B' $ states responsible for predissociation, causes severe level shifts~\cite{Namioka1964}: the $\Lambda$-doubling splitting even in the bound region below the $n=2$ dissociation limit. Predissociation of the $D^{1}\Pi_{u}$ state results in H($n=2$) excited atomic fragments\cite{Guyon1979}. In the case of production of H($2p$) the dissociation product can be observed via Lyman-$\alpha$ fluorescence and the H$(2s)$ atoms are observed in most cases via the same fluorescence due to collisions to the H($2p$) state.


The physics of a bound state interacting with a continuum state, where both states are simultaneously excited from a ground level was described by Fano\cite{Fano1961}.
If there is oscillator strength in both channels such interference leads to an asymmetric line shape, also referred to as a Beutler-Fano line shape, which is described by a $q$-parameter, indicating the degree of asymmetry \\
  \begin{eqnarray}
   q = \frac{<\varphi_{Dv'}|T_{DX}(R)|\varphi_{Xv''}>}{\pi V<\varphi_{B'\epsilon}|T_{B'X}(R)|\varphi_{Xv''}>}
\label{Qfactor}
\end{eqnarray}
In this formula the nominator stands for the transition dipole matrix element ($T(R)$ being
the $R$-dependent transition dipole moment) for excitation from the ground state $X^1\Sigma_g^+$ to the
excited state, in the present case a certain ro-vibrational level in the $D^1\Pi_u$ state \cite{Glass-Maujean1989}.
The denominator contains the matrix element for excitation to the continuum channel, which is in the present specific case represented by the $B'^1\Sigma_u^+$ state. $V$ represents the interaction matrix element between the two excited channels: discrete and continuum, in this case the $D$ and $B'$ states signifying the coupling of the bound state with the continuum. The matrix element for the rotational coupling operator, $H$, yields for the interaction matrix element~\cite{Glass-Maujean1979,Hougen1970} 
 \\
\begin{eqnarray}
 V= \langle\varphi_{B'\varepsilon}|H|\varphi_{Dv'}\rangle \propto \sqrt{J(J+1)}
\label{Vfactor}
\end{eqnarray}
hence the Fano $q$ parameter is inversely proportional to rotation. The predissociation widths $\Gamma$ are related to the square of  the interaction matrix element
\begin{eqnarray}
\Gamma=2\pi V^{2}
\label{eqn:Width}
\end{eqnarray}
also exhibiting a rotational dependence proportional to $J(J+1)$.

Various theoretical calculations of the widths of the $\Pi^+$ parity components of the $D^{1}\Pi_{u}$ states surfaced in the early seventies by Julienne\cite{Julienne1971} and Fiquet-Fayard and Gallais\cite{Fiquet-Fayard1971}, showing that the predissociation in the $D^1\Pi_u^+$ state must indeed be attributed to an interaction with the $B'^1\Sigma^+_u$ continuum.
Initially the two calculations were in disagreement with one another until a missing factor of 4 was discovered in the calculation by Julienne \cite{Fiquet-Fayard1972}. The two theoretical studies were still in disagreement with the measured values of Comes and Shumpe\cite{Comes1971} by roughly 25 percent. The experimental widths varied in a way not predicted by theory. Further measurements using the $J$ dependence of $q$, $\Gamma$ and of the intensities by Glass-Maujean \etal \cite{Glass-Maujean1979} gave a far better agreement with Julienne's \cite{Julienne1971} and Fiquet-Fayard and Gallais's \cite{Fiquet-Fayard1971} values for $v=3-5$ but still disagreed by roughly 25$\%$ for $v=7-11$.

\begin{figure}
\begin{center}
\includegraphics[width=1\linewidth]{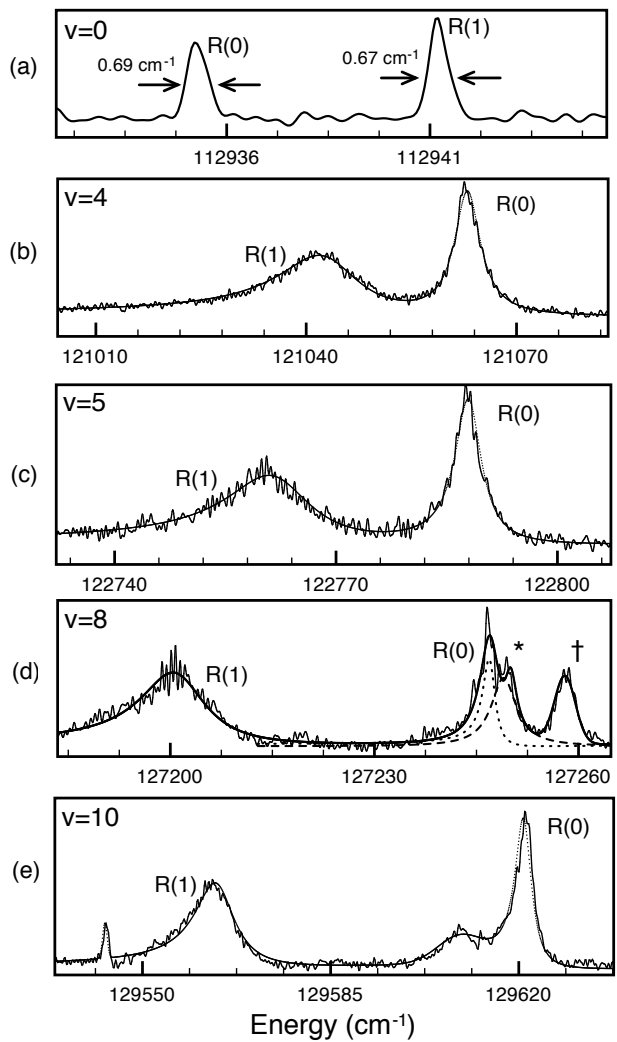}
\caption[]{Detail spectral recording of the $D-X\,(v,0)$ bands for $v = 0,4,5,8$ and $10$. The (0,0) band is below the second dissociation limit and is not predissociated; this serves to illustrate the limiting resolution of the spectrometer. The R(0) transition of the (8,0) band is blended by the R(0) transition from the $6p\sigma $ $v=3$ Rydberg series marked with a star. The line marked with a $\dagger$ is $9p\sigma$ $v=2$ R(1) transition. For the $ v=4,5 $ and $ 10 $ bands the final representation of a Fano-profile as following from the deconvolution procedure is also shown. These are the functional forms represented by the $q$ and $\Gamma$ parameters as listed in Table. \ref{tab:NewH2} and Fig. \ref{fig:q-factors}.}.
\label{fig:FanoFits}
\end{center}
\end{figure}

The accuracy of the theoretical modelling of the predissociative widths is determined mainly by the accuracy of the potential curves used.  The coupling to the $B$ and $C$ states does not have an effect on the predissociative widths but does play a role in the H$(2s)$/(H$(2s)$+H$(2p)$) branching ratio. The coupling to the $B$ state was included in the calculations by Borondo \etal \cite{Borondo1982} in their calculation of the branching ratio. Beswick and Glass-Maujean \cite{Beswick1988} conducted further studies on this topic including the predissociation of the $D$ state. Line shapes were obtained from solving the appropriate coupled Schr\"{o}dinger equation, yielding values for the predissociative widths and $q$-parameters. In a study by Mruga\l{}a \cite{Mrugala1988} the shifts between $\Pi^{-}$ and $ \Pi^{+} $ levels were calculated. Mruga\l{}a benefited from the accurate \emph{ab initio} calculations of the $ D-B' $, $ D-B $, $D-C$ and $D-D'$ couplings together with improved potential energy curves for the $ B,B',C,D $ and $ D' $ states \cite{Wolniewicz1988}. All non-adiabatic interactions between the $ D $, $ B $ and $ B' $ state were included within the close-coupling approach.


A multi-channel-quantum-defect (MQDT) approach was followed by Gao \etal \cite{Gao1993} In the frame work used the interactions between the $ D,B' $ and $ C $ states were included in a noniterative eigenchannel R-matrix approach. The effect of ionization was thought to be small in comparison to predissociation and was not included in the calculation. Furthermore the $B$ state interaction was neglected because of weak coupling to the $ D $ state. The calculation yields widths for the $J=2$ level of the $ D $ state which closely match the experimental values observed by Glass-Maujean and co-workers \cite{Glass-Maujean1979}.

The $\Pi^-$ parity components of the $3p\pi D^{1}\Pi_{u}$ state, probed in Q-transitions, undergo only very weak predissociation, for which coupling to the lower lying $2p\pi\,C^{1}\Pi_{u}$ is the only symmetry-allowed possibility. The effect of this homogeneous perturbation in terms of predissocation widths of the Q(1) transitions was calculated by Glass-Maujean \etal~\cite{Glass-Maujean2010}. An MQDT analysis produced line positions and intensities for the Q$(J)$ $(J=1-4)$ absorption transitions~\cite{Glass-Maujean2009}. In MQDT calculations all the interactions of the $np\pi^{1}\Pi^{-}_{u}$ states were included and not only the $C$ and $D'$ states. The results of these calculations were consistent with observations of Lyman-$\alpha$ fluorescence and visible molecular fluorescence\cite{Glass-Maujean2010,Glass-Maujean2009}.

\section{Experimental Results and Discussion}

The vacuum ultraviolet Fourier-Transform spectra show the features of the $D^{1}\Pi_{u}-X^1\Sigma_g^+\,(v,0)$ bands for H$_{2}$ up to $v=17$, which is the uppermost bound vibrational level in this potential. Fig.~\ref{fig:FanoFits} displays detail recordings of some of
the regions with pronounced $D-X$ features. In order to retrieve these spectra the bell-shaped background continuum of the undulator profiles as shown in Fig.~\ref{fig:Bell-curve}, were transformed into a flat continuum by fitting the background with a Gaussian function. The measurement window was divided through by the Gaussian fit resulting in a flat background. In a second step the Beer-Lambert absorption depth was linearized, and the scale inverted to arrive at the spectra depicted in Fig.~\ref{fig:FanoFits}.  The spectrum of Fig.~\ref{fig:FanoFits}(a) shows the unpredissociated $D-X\,(0,0)$ band; hence these lines represent the limiting resolution of the spectral method. The widths of $\sim 0.7$ \wn\ are a result of Doppler broadening (0.6 \wn\ for H$_2$ at 100 K) and the instrumental width of 0.35 \wn\ from the settings of the FT-instrument.

\begin{figure}
\begin{center}
\includegraphics[width=1\linewidth]{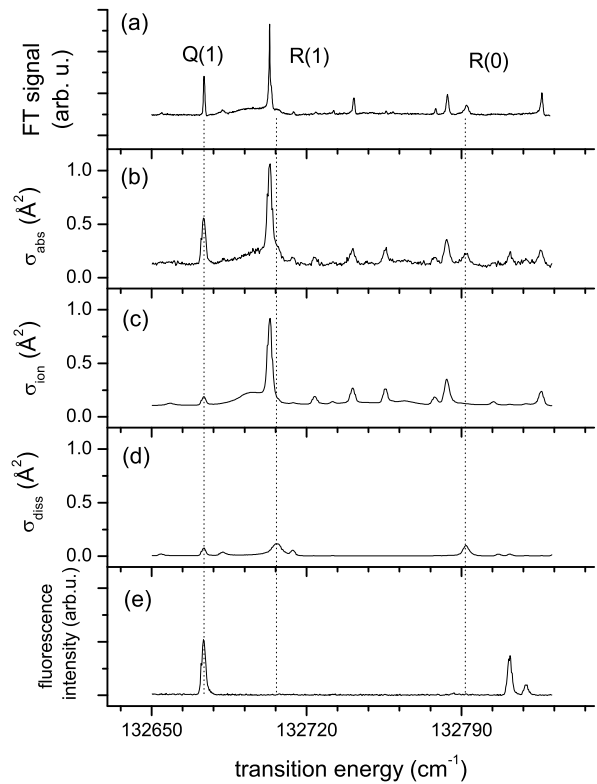}
\caption[]{The $D-X\,(14,0)$ band of H$_{2}$ displaying the R(0), R(1)
and Q(1) lines recorded by various methods.
(a) The high resolution absorption spectrum recorded with the Fourier-transform setup
at SOLEIL; (b) The absorption spectrum recorded with the 10 m normal incidence scanning
monochromator at room temperature; (c) The photoionization spectrum; (d) The spectrum recorded
by detection of Lyman-$\alpha$ photons originating from the dissociation fragments;
(e) The fluorescence from the H$_2$ $D$ state decaying to high-lying levels of $g$
symmetry in the molecule, with cascades from these $g$ states to the $B$ state~\cite{Glass-Maujean1985}. Note that the intensities in spectra (b-d) represent absolute
cross sections. Spectra (b-e) were obtained at BESSY.}
\label{fig:H2v=14}
\end{center}
\end{figure}

The uncertainties in the line positions are governed by the signal to noise ratio, the width of the transition and the number of points that the transition consists of~\cite{bookDavis2001}. When all systematic effects \cite{Ivanov2010} of the FT spectrometer are taken into account the accuracy is limited to 0.03 \wn. The so-called $p$ parameter \cite{deOliveira2009}, connected to an interferometric setup with a HeNe reference laser controlling the path length travel, is adjusted depending on the operational wavelength regime. It was discovered that the changing of this parameter induces small systematic effects on the wavelength calibration \cite{Ivanov2010}. In the present study these effects were not accounted for and so a conservative estimate of the statistical deviation in the line positions for the  unpredissociated lines ($ v=0-2 $) and the $ \Pi^{-} $ components is 0.06 \wn.
For the broad $ \Pi^{+} $ components the uncertainty is in the range of 0.3 \wn, while for the blended bands (7,0), (8,0) and (9,0) the uncertainty increases to 0.6 \wn.

\subsection{Spectroscopy}

The transition energies resulting from the combined set of spectra are listed in Table~\ref{tab:NewH2}. The corresponding level energies are available via the EPAPS data depository of the American Institute of Physics~\cite{EPAPS}. Various line shape fitting methods were employed to derive the line positions; for the asymmetric Fano line shapes the true transition frequency was derived from the deconvolution procedure explained in Subsection C.

The assignment of the unpredissociated lines derives straightforwardly from the accurate emission data\cite{Abgrall1994,Bailly2010}, while for higher lying levels for the $D^1\Pi_u^-$ components the multichannel quantum defect calculations\cite{Glass-Maujean2009} are a guidance. For the transitions to the  $^{1}\Pi^{+}$ components of $v=14-17$ that information is lacking. These levels were identified starting with the  $^{1}\Pi^{-}$ components \cite{Abgrall1994,Glass-Maujean2009} and setting the $\Lambda$-doubling to zero. Where lines appear to be blended the assignments and analysis was aided by the dissociation spectra. In particular the R(1) line of the (14,0) band was obscured by a particularly strong, unidentified transition. As shown in Fig.~\ref{fig:H2v=14} this feature could not have been assigned properly without the aid of the dissociation spectrum.

Other lines analyzed in the dissociation spectrum are the R(0) and R(1) lines in the $(16,0)$  band and the R(1) line in the $(17,0)$ band. These transitions were rather weak in the FT-absorption spectrum but appeared clearly in the dissociation spectrum. Due to the lower resolution of the spectrometer at BESSY and to the higher temperature of the gas, the uncertainty of the position for these measurements is slightly better than $1$ \wn.

 Several additional problems determining accurate line positions and line parameters occurred. The R(1) transition  in the $D-X$ (6,0) band is blended by a transition to the $B''$ state. This presents a complication in that the lines are blended in both the absorption and dissociation spectra since both states predissociate. In this case the blending with the R(0) transition in the $B''-X$ (4,0) band is such that no data could be extracted. Similarly the R(1) transition in the $D-X$ (7,0) band is blended by the R(1) transition in the $B''-X$ (5,0) band~\cite{Glass-Maujean2007a} and by the Q(1) transition of the $D'-X$ (4,0) band~\cite{Takezawa1970}; no data on the predissociated widths could be extracted. The position specified in Table~\ref{tab:NewH2} are those from the deconvoluted fit. Due to the relatively small shift of the line position~\cite{Gao1993} -2.7 \wn\ (from the interaction with the continuum), we believe that the fit does provide the true line centre. The R(0) transition of the (7,0) band is blended by a R(0) transition from the $5p\sigma$ $v=3$ Rydberg series~\cite{Herzberg1972}, position and widths were determined by deconvolving with the overlapping line. The R(0) transition in the (8,0) band is blended with an R(0) transition from the $ 6p\sigma $ $(v=3)$ Rydberg series\cite{Dehmer1976}. Again a fit was achieved from which an overlapping line was deconvolved, this is shown in Fig. \ref{fig:FanoFits}(d). The (9,0) band was superimposed upon three different Rydberg series~\cite{Dehmer1976} ($ 17p\pi $ $ v=2 $, $ 22p\sigma $ $ v=2 $, $ 7p\pi $ $ v=3 $); line positions and widths for R(0) and R(1) were determined by deconvolution.

The transitions of the first three unpredissociated bands $v=0-2$ were compared to the measurements of Bailly~\etal~\cite{Bailly2010} which are accurate to $0.005$ \wn\ for the $ D $ system. The comparison is good with an average deviation of $-0.03$ \wn\ thus demonstrating the high absolute accuracy of the FT instrument. The fitted positions of the $ \Pi^{-} $ states were compared with the measurements of Abgrall~\etal~\cite{Abgrall1994} for $v=3-14$ which we consider to be accurate to within 0.1 \wn. The overall agreement between these Q lines is better than $0.1$ \wn\ except for $v=14$.

We compare to the measurements of Glass-Maujean \etal~\cite{Glass-Maujean2007} for the $ \Pi^{-} $ components probing Q(1) transitions for $v=15-17$ in Table \ref{tab:NewH2}. There appears to be a discrepancy between our measurements and those compiled by Glass-Maujean \etal \cite{Glass-Maujean2007} for the last three bands, (15,0) (16,0) and (17,0) yet this is still within the experimental uncertainty of 6 \wn for that specific experimental configuration. In a later publication these measurements were compared to a MQDT calculation \cite{Glass-Maujean2009} and discrepancies in the order of 5 \wn\ were found for these three bands once again within the experimental uncertainty. The present measurements side with the MQDT calculations and the comparison can be found in Table~\ref{tab:NewH2}. The comparison is suggestive of an accuracy better than 1 \wn\ thus confirming the high accuracy of these calculations. 

For the $\Pi^+$ components the present measurements were compared to those of Takezawa \cite{Takezawa1970} for $v=3-11$. Deviations amount to several 0.1 \wn\ due to the lower accuracy in the classical absorption study, where no Fano analysis was conducted and no pressure corrections were made \cite{Takezawa1970}. One exception was the R(0) transition in the (7,0) band where a deviation of more than 11 \wn\ was found, we attribute this to either a misassignment or a typo and therefore compare to the data of Herzberg and Jungen~\cite{Herzberg1972}. Lines in bands (12,0) and (13,0) were compared to the measurements of Coman and McCormack \cite{Croman2008}. The discrepancies of order \wn\ indicate that these laser-based data were not calibrated on an absolute scale.

\subsection{$\Lambda$-Doubling}

\begin{figure}[pt]
\begin{center}
\includegraphics[width=1\linewidth]{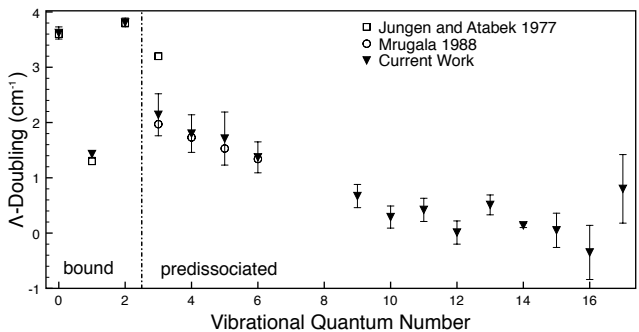}
\caption[]{Values of the $ \Lambda $ doubling for the $ D^{1}\Pi_{u} $ \, $ J=1$ levels as a function of vibrational quantum number. The sign is such that $ \Pi^{+} $ or $ \Pi_{(e)} $ levels are higher than $ \Pi^{-} $ or $ \Pi_{(f)} $ levels. The values of the $\Lambda$ doubling for $v=7$ and 8 have been omitted due to blending of the R(0) and Q(1) transitions.}
\label{fig:LambdaDoubling}
\end{center}
\end{figure}

In Fig. \ref{fig:LambdaDoubling} the splitting between different $ \Pi^{+} $ and $ \Pi^{-} $ parity levels for the $ J=1 $ level have been plotted as detected from the current measurements. This involves adding the ground state energy difference between rotational levels $J=0$ and $1$ to the Q(1) transitions and subtracting this from the R(0) transitions (the ground state rotational energy splitting of 118.486 \wn\ was taken from the accurate calculations of Wolniewicz\cite{Wolniewicz1995}). Values for this $ \Lambda $-doubling have been plotted for each vibration up to $v=17$. The two "accidents" $v=13$ and $v=17$ could be due to local couplings.

The lowest three bound states of the $ D $ system are subject to strong perturbations from the last few bound levels of the $ B' $ system \cite{Namioka1964,Abgrall1994}. This explains the "accidental" behavior seen for the $ \Lambda $-doubling in $v=1$ and $2$
in Fig.~\ref{fig:LambdaDoubling} and the reversal between the R(0) and R(1) lines in 
Fig.~\ref{fig:FanoFits}(a). The $\Lambda$-doubling turns into a smooth decaying function 
from $v\geq3$ toward high $v$. We postulate that the $\Lambda$-doubling is due to the interaction between the discrete $D$ states 
and the summed contribution of low-lying discrete levels in the $B'$ state and the $B'$ 
continuum. Hence, the physical origin of the $\Lambda$-doubling is connected to 
that of the origin of the predissociation of the $D$-state (discussed below). 
Both phenomena become weaker toward high vibrational levels, due to reduced 
vibrational wave function overlap between $ D $ and $ B'$ levels. 
The present data shows excellent agreement with calculations for the 
$ \Lambda $-doubling for vibrational levels~\cite{Jungen1977,Mrugala1988} $v=0-6$.
\begin{figure}[pb]
\begin{center}
\includegraphics[width=1\linewidth]{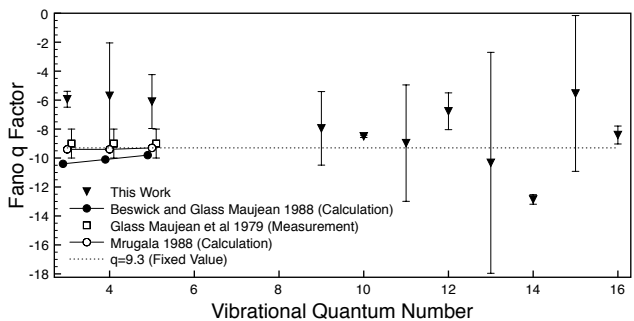}
\caption[]{The fitted $q$ parameters for the R(1) transitions compared with the measured values of Glass-Maujean \etal \cite{Glass-Maujean1979} and the calculated values of Beswick and Glass-Maujean \cite{Beswick1988} and Mruga\l{}a \cite{Mrugala1988}. Overlapping values have been slightly offset along the x-axis to make the error bars clearly visible.}
\label{fig:q-factors}
\end{center}
\end{figure}

\subsection{Predissociation analysis}

Initially a two component fit of the data in $ q $ and $ \Gamma $ was made with a Fano function. The contribution of the instrument and Doppler width (Gaussian $ \sim $0.7 \wn) was deconvolved from the observed line shapes with a triangular. In principle a Fano convolved Gaussian is warranted but there is no closed expression for this convolution and it has been demonstrated that the difference between the closed expression for a Fano convolved triangular and the numerical solution for a Fano convoluted Gaussian\cite{Jimenez-Mier1994} is less than 4$\%$.

\begin{figure}[pt]
\begin{center}
\includegraphics[width=1\linewidth]{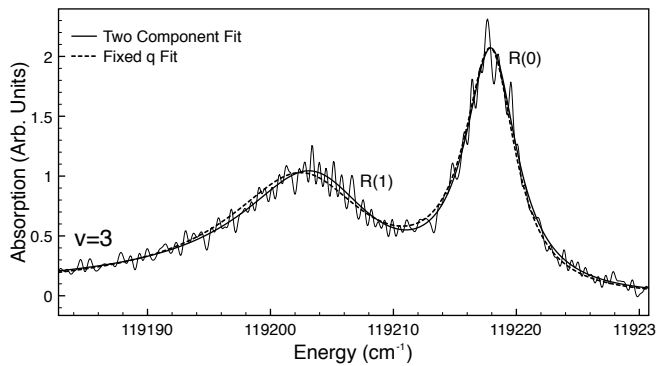}
\caption[]{FT-spectrum of the $ D-X $ (3,0) band of H$ _{2}$ with a comparison between the two component fit of $q$ and $\Gamma$ and a one component fit of $\Gamma$ only. The resulting values of the parameters for the two component fit were: $q=-5.90 \pm 0.55$ and $\Gamma=12.7 \pm 1.1$ \wn\ for R(1), $q=-43 \pm 35$ and $ \Gamma=5.10 \pm 0.26 $ \wn\ for R(0). In a one component fit fixing $q$ parameters to those obtained by theory \cite{Mrugala1988} yields: $\Gamma =13.6 \pm 1.0$ \wn\ for R(1) and $\Gamma=4.97 \pm 0.26$ \wn\ for R(0).}
\label{fig:FixedqFactors}
\end{center}
\end{figure}

The extracted $q$ parameters for the R(1) transitions are shown in Fig. \ref{fig:q-factors}. Determination of the $q$ parameter is complicated due to two reasons: Firstly the noise on the spectrum and secondly imparting a resolution condition on two superimposed Fano functions. In Fig. \ref{fig:FixedqFactors} this is illustrated for the specific case of the $ D-X $ (3,0) band, where a rather accurate value of $q =-5.90 \pm 0.55$ is derived in a two component fit optimizing for $q$ and $\Gamma$. In the figure we compare the two component fit to a one component fit of $\Gamma$ only. The fit is insensitive to a variation in $q$, while there is very little correlation between the two parameters. This is most clear for the R(0) transition where the $q$ parameter has doubled yet the width changes by just over 0.1 \wn. Due to the broad, asymmetric nature it is difficult to address  how far these lines are resolved. This may lead to an erroneous determination of the $ q $ parameter. To compare this to the ideal case, the R(1) transition for $v=10$ is clearly resolved with good signal to noise ratio (see Fig. \ref{fig:FanoFits}(e)). The resulting $q$ parameter determined is $-8.50 \pm 0.08$ which compares well to the calculated value expected to be close to $-9$.  In order to obtain the $\Gamma$ parameters specified in Table \ref{tab:NewH2} the $q$ parameters were fixed for the R(0) and R(1) transitions to the calculated $v$ dependent values \cite{Mrugala1988} for $v=3-5$. For the remaining vibrations $v=8-17$ values of $-19.55$ for R(0) transitions and $-9.3$ for the R(1) transitions were used~\cite{Mrugala1988}.


\begin{figure}[pb]
\begin{center}
\includegraphics[width=1\linewidth]{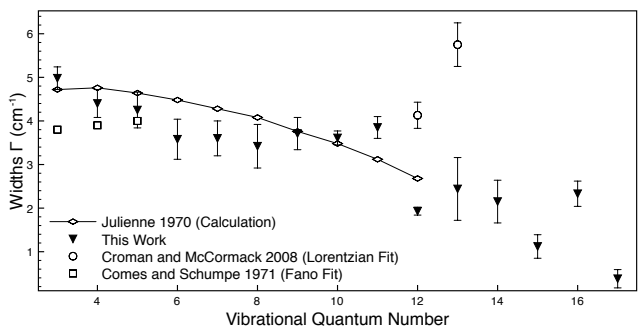}
\caption[]{The deduced values of the predissociated widths $ \Gamma $ for the $J=1$ level compared with other measured and calculated values currently available in the literature. The calculations of Julienne \cite{Julienne1971} have been multiplied by a factor of 4 to correct for an error as specified by Fiquet-Fayard and Gallais \cite{Fiquet-Fayard1972}.}
\label{fig:WidthsJ=1}
\end{center}
\end{figure}

\begin{figure}[t]
\begin{center}
\includegraphics[width=1\linewidth]{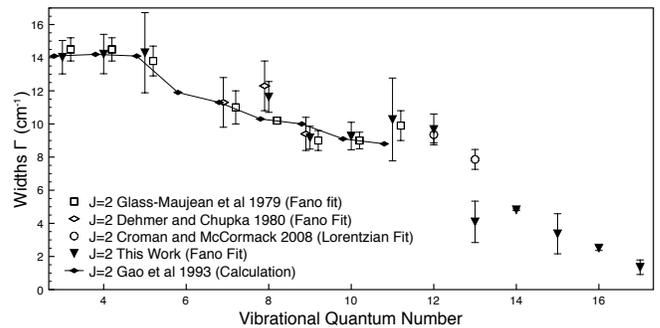}
\caption[]{The deduced values of the predissociated widths $ \Gamma $ for the $J=2$ level compared with other measured and calculated values currently available in the literature. Overlapping values have been slightly offset along the x-axis to make the error bars clearly visible.}
\label{fig:WidthsJ=2}
\end{center}
\end{figure}

The results for the widths of the $J=1$ levels are displayed in Fig.~\ref{fig:WidthsJ=1}. To our knowledge there is no MQDT calculation for $ J=1 $ levels and measured values are lacking still. The model of Julienne \cite{Julienne1971}, after applying the necessary corrections~\cite{Fiquet-Fayard1971}, reproduces the overall decrease in the predissociation widths, but fails to cover some of the details, although deviations are not significant in all cases. It is the first time that the widths of the $J=1$ level series are determined, except for $v=12$ and 13 determined previously by Croman and McCormack \cite{Croman2008}. The present data are in agreement with the calculations of Julienne \cite{Julienne1971}, after multiplication by a factor of 4, for $v=3-5$, and for higher $v$ values in much better agreement than those of Croman and McCormack \cite{Croman2008}.


The present data in the predissociated linewidth parameter $ \Gamma $ for $ J=2 $ are displayed in Fig.~\ref{fig:WidthsJ=2} and are in good agreement with previous observations, in particular with those of Glass-Maujean \etal \cite{Glass-Maujean1979} extending to $v=11$ and Dehmer and Chupka \cite{Dehmer1980} for $v=8$ and $9$. The value at $v=12$ is in agreement with that of Croman and McCormack \cite{Croman2008}, while the value for $v=13$ deviates, similarly as both values for $J=1$ in Fig. \ref{fig:WidthsJ=1}. Apart from the latter the present data and most reliable previous data produce an overall trend of smoothly decreasing predissociating widths $\Gamma$ toward high vibrational levels. This trend is reproduced by calculations in different frameworks \cite{Julienne1971,Gao1993}, and is also consistent with observations on the $\Lambda$-doubling (see Fig.~\ref{fig:LambdaDoubling}).

The data suggest a slight increase of predissociation at $ v=11 $, an effect that was hypothesized by  Glass-Maujean \etal \cite{Glass-Maujean1979} and later by Gao \etal \cite{Gao1993} as resulting from a contribution by an additional predissociation channel, the $ B''\bar{B} $ double well potential \cite{deLanga2001} intersecting the $ D $ state near 5.5 a.u. as shown in Fig.~\ref{fig:PotentialEnergyCurves}. It is noted that the $ B''\bar{B} $ state is bound at this energy but predissociates due to the same $ B' $ continuum. Interference between the two dissociation paths may affect the rate of predissociation in the $ D $ state. 
Future MQDT calculations may elucidate whether this phenomenon indeed occurs at $ v=11 $. Also such calculations may explain the observed values for $ \Gamma $ in the range $ v=13-17 $.

\section{Conclusion}

The superior resolution of the novel XUV Fourier-transform instrument is
applied for a reinvestigation of the $D-X$ absorption system in H$_2$. This
system is a benchmark system for the study of predissociation in diatomic
molecules with the special feature of pronounced Fano-type line shapes. In
H$_2$, the smallest neutral molecular system, the widths and $q$-asymmetry
parameters can be calculated in various first principles schemes to be
compared at a high level of accuracy to experiment. The presently measured
transition frequencies are the most accurate and signify an order of
magnitude improvement in accuracy over previous studies \cite{Monfils1965,Takezawa1970,Comes1971}. The data
are extended up to the highest $v=17$ vibrational level in the $D^1\Pi_u$
state. The FT spectrometer has a far better resolution than that obtained by scanning monochromators but with the measurement limited to direct absorption only it cannot provide all the information of the interaction between light and molecules. In this way the combination of the two experimental techniques, the FT absorption spectrum from SOLEIL and the scanning monochromator measurements from BESSYII, are complimentary. This leads to an almost complete characterization of the molecular state.

The extracted predissociation widths are found to decrease up to
$v=10$ as expected from theory, with indication of a possible sudden
increase in $\Gamma$ at $v=11$, which might be explained by the opening up of an
additional predissociation channel associated with the $B''\bar{B}$ double
well state. The smooth development of a decreasing $\Lambda$-doubling in
the $D^1\Pi_u$ state toward higher vibrational levels is consistent with
the decrease in the predissociation rates and in excellent agreement with
calculation~\cite{Jungen1977,Mrugala1988} for $v=0-6$; both phenomena have their origin in overlap
with the $B'^{1}\Sigma_{u}^{+}$ state. The positions of the $D^{1}\Pi_{u}^{-}$
levels are measured to an accuracy of 0.06 \wn, which is the most accurate
to date. These data compare extremely well to previously calculated positions within the
MQDT-framework~\cite{Glass-Maujean2009} suggesting an accuracy better than 1 \wn\ and providing further
proof for its suitability for the modelling of such complex molecular
processes at very high excitation energy in the molecule.

\section*{Acknowledgement}

The staff of SOLEIL and its DESIRS beamline is thanked for the support and for providing the opportunity to conduct measurements in campaigns in 2009. The work was supported by the Netherlands Foundation for Fundamental Research of Matter (FOM). We are indebted to EU for its financial support via the Transnational Access funding scheme. MGM acknowledges help from P. Reiss, H. Schmoranzer and the BESSY staff, and is indebted to EU support (grant ELISA n$^{o}$ 226716). WLTB is indebted to support from the French ANR project SUMOSTAI.

\clearpage
\renewcommand{\thefootnote}{\alph{footnote}}
\begingroup
\squeezetable
\begin{longtable}{l l r r r r }
 \caption{Table of transition energies in \wn\ and predissociated linewidths $\Gamma$ in \wn\  for the measured transitions in the $D^{1}\Pi_{u}-X^{1}\Sigma^{+}_{g}$ $ (v,0) $ system of H$ _{2} $. The widths measured for the unpredissociated levels $v=0-2$ and the Q(1) transitions are limited by the Doppler broadening and therefore not specified in the table. $\Delta$ represents deviations between present values minus previously published data. For excited states $v=0-2$ comparison was made with Bailly \etal \cite{Bailly2010} For  the $ \Pi^{+} $ components in $v=3-11$ comparison was made with Takezawa \cite{Takezawa1970}. For  $v=12$ and 13 we compare to Croman and McCormack \cite{Croman2008} and for the remaining bands the measurements of Soleil were compared to those from BESSY. The $ \Pi^{-} $ components of $v=3-14$ are compared with the measurements of Abgrall \etal \cite{Abgrall1994} and for $v=15-17$ we compare to Glass Maujean \etal \cite{Glass-Maujean2007} $\Delta^{M}$ represents a comparison with the MQDT calculations of Glass-Maujean and Jungen \cite{Glass-Maujean2009}; these calculations are available for the Q(1) transitions only.}
\label{tab:NewH2}
\\
\colrule
\multicolumn{2}{c}{Transition Energy} & \multicolumn{1}{c}{$\Delta$}  & \multicolumn{1}{c}{$\Gamma$}  & \multicolumn{1}{r}{$\Delta^{M}$} \\
\colrule
\endfirsthead
\\
\colrule
\multicolumn{2}{c}{Transition Energy} & \multicolumn{1}{c}{$\Delta$}  & \multicolumn{1}{c}{$\Gamma$} & \multicolumn{1}{r}{$\Delta^{M}$}   \\
\colrule
\endhead


 $D-X (0,0)$ & & & & \\
 \\Q(1)&112\,813.14 &-0.03&---&-0.39&\\
     R(0) &112\,935.25& -0.03 & ---  &---  & \\
     R(1) &  112\,941.20  & 0.03 & --- &---  & \\

 \\
  $D-X (1,0)$ & & & &  \\
 \\Q(1)&115\,035.88 &-0.04&---&-0.54&\\
     R(0) &  115\,155.80  & -0.01 & --- &---  &\\
     R(1) &  115\,151.14  & -0.02 & --- &---  &\\
 \\
 $D-X (2,0)$ & & & &  \\
 \\Q(1)&117\,129.34&-0.04&---&-0.67&\\
     R(0) &  117\,251.64  & -0.07 & ---& ---  &\\
     R(1) &  117\,244.82  & -0.02 & ---&---  &\\
 \\
 $D-X (3,0)$ & & & &  \\
 \\ Q(1)\footnotemark[19]&119\,097.37&0.02&---&-0.65&\\
     R(0) &  119\,217.99  & 0.39 & 4.94  &---  &\\
     R(1) &  119\,203.60  &1.6&13.72&---  &\\
 \\
 $D-X (4,0)$ & & & & \\
 \\Q(1)\footnotemark[19]&120\,942.76&0.05&---&-0.7&\\
     R(0) &  121\,063.05  & 0.65 & 4.40&---  & \\
     R(1) &  121\,042.48  & 0.48 &14.22&---  & \\
 \\
 $D-X (5,0)$ & & & &  \\
 \\Q(1)\footnotemark[19]&122\,667.75 &0.06&---&-0.76&\\
     R(0) &  122\,787.94  & 0.54&4.25&---  &\\
     R(1) &  122\,760.86  & 0.16& 14.30& ---  &\\
 \\
 $D-X (6,0)$ & & & &  \\
 \\Q(1)&124\,273.90 & 0.03 & ---&-0.71&\\
     R(0) &  124\,393.75  & 0.15&3.58  &  ---  &\\
     R(1)\footnotemark[2] &  ---  & --- & --- &---  &  \\
 \\
 $D-X (7,0)$ & & & &  \\
 \\Q(1)\footnotemark[2]&125\,759.85 &0.03 & ---&-0.59&\\
     R(0)\footnotemark[2]\footnotemark[1]&  125\,877.03  & 0.43&3.60& ---  &\\
     R(1)\footnotemark[2]&  125\,841.35  & 0.55 &---&---  &\\
 \\
 $D-X (8,0)$ & & & &  \\
 \\Q(1)&127\,129.23& 0.09 &---& -0.57\\
     R(0)\footnotemark[2]&  127\,246.84 & -1.66 &3.42&---  &\\
     R(1)&  127\,201.15  & -1.35 & 11.63&---  &\\
 \\
 $D-X (9,0)$ & & & &  \\
 \\Q(1)&128\,377.38 &-0.07&---&-0.44&\\
     R(0)\footnotemark[2]&  128\,496.54  & 0.34 & 3.71&---  &  \\
     R(1)\footnotemark[2]&  128\,444.94  & 0.04 &9.17&---  & \\
 \\
 $D-X (10,0)$ & & & & \\
 \\Q(1)&129\,502.50 &0.13&---&-0.21&\\
     R(0) &  129\,621.43  & -0.07&3.61&---  & \\
     R(1) &  129\,563.49  & 0.79&9.27& ---  & \\
 \\
 $D-X (11,0)$ & & & & \\
 \\Q(1)&130\,499.61&0.09&---&0.05&\\
     R(0) & 130\,618.52&-3.18&3.85&---  &\\
     R(1) &  130\,554.15 & -0.45& 10.27  &---  & \\
 \\
 $D-X (12,0)$ & & & &  \\
 \\Q(1)&131\,366.00&0.10& ---&0.21&\\
     R(0) &  131\,484.50  & -0.77& 1.93&---  & \\
     R(1) &  131\,414.00  & -0.76 & 9.67&---  &\\
 \\
 $D-X (13,0)$ & & & & \\
 \\Q(1)&132\,093.33 &-0.11&---&-0.29&\\
     R(0) &  132\,212.33  & 1.96  &2.44&---  &\\
     R(1) &  132\,134.47  & -2.04 &4.09&---  &\\
 \\
 $D-X (14,0)$ & & & & \\
 \\Q(1)&132\,673.74 &0.49&---&-0.67&\\
     R(0) &  132\,792.37  & -0.76&2.15&---  &\\
     R(1)\footnotemark[2]\footnotemark[4] &  132\,706.89  &---& 4.82&---  & \\
 \\
 $D-X (15,0)$ & & & & \\
 \\Q(1)&133\,100.81 &-7.13&---&-0.78&\\
     R(0) &  133\,219.34  & 0.49 & 1.12  &---  & \\
     R(1) &  133\,124.96  & 0.17 & 3.30 &---  & \\
 \\
 $D-X (16,0)$ & & & &  \\
 \\Q(1)&133\,366.45 &-6.01&---&-0.67&\\
     R(0)\footnotemark[4] &  133\,484.59  & --- & 2.33&---  & \\
     R(1)\footnotemark[2]\footnotemark[4] &  133\,381.11  & ---& 2.51 &---  &\\
 \\
 $D-X (17,0)$ & & & &  \\
 \\Q(1)&133\,468.31 & -5.44& ---&0.17&\\
     R(0) &  133\,587.77  &-0.27 &0.38&---  & \\
     R(1)\footnotemark[4] &  133\,472.62  &---&1.35&---  &\\
  \\

 \footnotetext[1]{Compared with the data of Herzberg and Jungen \cite{Herzberg1972}}
 \footnotetext[2]{Blended in the absorption spectra}
 \footnotetext[19]{Slightly saturated}
 \footnotetext[4]{Data from dissociation spectra}

\end{longtable}
\endgroup


\clearpage
\include{Table1.tex}
\clearpage

\end{document}